\documentclass[11pt,showpacs]{revtex4-1}

\usepackage{epsfig}
\usepackage{graphicx}
\usepackage{amssymb}
\usepackage{amsmath}
\usepackage{amsthm}
\usepackage{amsopn}

\def\comment#1{}

\def\beq{\begin{equation}}
\def\eeq{\end{equation}}
\def\bea{\begin{eqnarray}}
\def\eea{\end{eqnarray}}

\def\ep{\epsilon\!\!\!/}

\usepackage{ulem}
\usepackage{cancel}
\usepackage{color}

\begin{document}

\title{ Evidence for cosmic neutrino background form CMB circular polarization }

\author{Rohoollah Mohammadi \footnote{rmohammadi@ipm.ir}}
\affiliation{ School of physics, Institute for research in fundamental sciences (IPM), Tehran, Iran.\\
Iran Science and Technology Museum   (IRSTM), PO BOX: 11369-14611, Tehran, Iran.}

\begin{abstract}
The primordial anisotropies of the cosmic microwave background are linearly polarized via
Compton-scattering. On the other hand, a primordial degree of circular polarization of the Cosmic Microwave Background is
not observationally excluded. In this work, we discuss the generation of the circular polarization of CMB via their scattering on the
cosmic neutrino background since the
epoch of recombination.  We show that photon-neutrino interaction can transform plane polarization into circular polarization through processes $\gamma+\nu\rightarrow\gamma+\nu$ and the Stokes-V
parameter of CMB has linear dependence on the wavelength and the cosmic neutrino background perturbations of pressure and shear stress and also  the maximum value of $C^V$ is estimated in range of a few Nano-Kelvin square.
\end{abstract}

\maketitle
\section{ Introduction}
Modern cosmological observations of the Cosmic Microwave
Background (CMB) radiation  contain valuable information about our universe.
The CMB photons have decoupled from matter about $3\times 10^5$ years after the Big-Bang (BB), so  we are unable to probe the universe closer than 300 000 years to the BB by using CMB.
Cosmological information encoded in the CMB radiation concerns not only
temperature fluctuations and the spectrum of anisotropy pattern, but
also the intensity and spectrum of linear and circular
polarizations. From a result of the anisotropic Compton scattering around the
epoch of recombination, it is generally expected that some relevant linear polarizations (about $10\%$) of CMB radiation should be present \cite{cosowsky1994,zal,hu},
and polarization fluctuations are smaller than the temperature
fluctuations  \cite{nature}.
Currently, there are several ongoing experiments \cite{exp1,exp2,exp3,exp4,exp5,exp6,plank}  attempting to
measure CMB polarizations.
Theoretical studies of CMB
polarizations were carried out in Refs.~\cite{cosowsky1994,zal,hu,kaiser1983}, and
numerical calculations \cite{num1,num2} have confirmed that about
$10\%$ of the CMB radiation is linearly polarized, via the
Compton and Thompson scattering of unpolarized photons at the
last scattering surface (the redshift $z\sim 10^3$).
 Polarized light is conventionally described in terms of the Stokes parameters, a linearly polarized radiation is described by non-zero values for the Stokes parameters Q and/or U
and the possibility of the generation of circular polarization can be determined by the Stokes parameter V \cite{jackson}. On the basis of the mechanism discussed in \cite{cosowsky1994}, the linear polarization of the CMB in the presence of a large-scale magnetic field $B$ can be converted to the circular polarization under the formalism of the generalized Faraday rotation (FR) \cite{faraday con,fc} known as the Faraday conversion (FC). The evolution of the Stokes parameter V given by this mechanism is obtained as
\bea
\dot{V}=2U\frac{d}{dt}(\Delta\phi_{FC})
\label{fc},
\eea
where $\Delta\phi_{FC}\propto B^2$ is the Faraday conversion phase shift \cite{fc}. There are several papers which have attempted to discuss the probability of the generation of circular polarization of CMB photons. Giovannini has
shown that if the CMB photons are scattered via electrons in the presence of a magnetic field, a non-vanishing
V mode can be produced \cite{gio1,gio2}.   Furthermore, Cooray, Melchiorri and
Silk have discussed that the CMB radiation observed today is not exactly the same as the
field last scattered \cite{fc},  Bavarsad $et$ $al$ have shown  that CMB polarization acquires a small degree of circular polarization when a background magnetic field is considered or the quantum
electrodynamic sector of standard model is extended by Lorentz non-invariant operators as well as
non-commutativity \cite{khodam}, Motie and Xue  have discussed that the circular polarizations of radiation fields can be generated
from the effective Euler-Heisenberg Lagrangian \cite{xue} and the transform plane
polarization into circular polarization via photon-photon interactions mediated by the neutral hydrogen background, $\gamma+\gamma+atom\rightarrow \gamma+\gamma+atom$, through completely forward processes, has been discussed by Sawyer \cite{pp}. We would like to point out that photon - neutrino scattering can generate circular polarization. The reason for that is:  in context of standard model we have the purely left-handed interaction for neutrinos which caused linearly polarized photons achieve circular polarizations by interacting with left-handed neutrinos, in contrast they do not acquire circular polarizations by interacting with electrons in the forward scattering terms of \cite{cosowsky1994}. We can consider any linear polarization as two equal component, lift and right handed circular polarization. Due to lift handed interaction of neutrino, only one part of this linear polarization (lift handed component) is affected by neutrino. Finally after neutrino - photon scattering, the total number of lift and right handed circular polarization become different and we have a net circular polarization for photons. In this study, we are going to check and calculate the generation of circular polarization for CMB due to their scattering with the cosmic neutrinos background  (C$\nu$B).

On the other hand, a similar probe like the CMB is the cosmic neutrinos background  (C$\nu$B) which can give us very helpful information about the early universe. Due to their weak
interaction they decouple earlier about 1 second after the BB from matter at a temperature of
$T_{\nu}\approx 1$MeV ($10^{10}$ Kelvin). The Cosmic Neutrino Background C$\nu$B of today (with temperature $T_{0\nu}\approx 1.95$ K)
therefore  contains information of the universe already 1 second after the BB. The CMB measurements  have already constrained the neutrino and antineutrino masses sum within LambdaCDM framework to about $\sum_{i=e,\mu,\tau}(m_{\nu_i}+m_{\bar\nu_i})\sim 0.6$ eV, implying a constraint of $\sim0.1$ eV on the individual neutrino and antineutrino mass. We consider this limit for mass of each type of neutrino and antineutrino for the rest.
 But we should remind that the detection of this C$\nu$B seems to be hardly possible due to the weak interaction of neutrinos with matter and
due to their low energy. Nevertheless, several methods have been
discussed in the literature to search for these relic neutrinos \cite{katrin,katrin1,katrin2,katrin3,katrin4,katrin5,katrin6,katrin7,katrin8,katrin9}. Here we discuss the possibility to find any effects of C$\nu$B on the circular polarization of the CMB photons via photon-neutrino scattering. As is well known, photon-neutrino cross-section in the context of standard model  is very small because neutrinos are neutral particles with very small electromagnetic dipole moment $\mu_\nu\propto m_\nu$ and also  the leading order of photon-neutrino interaction (one-loop) contain the weak interaction. But this is not really bad news for our idea because we are going to consider the last scattering surface for photon-neutrinos around the epoch of recombination. This means if coherent photon-neutrino forward scattering after recombination age can provide any sources for the circular polarization of CMB, $\Delta\phi_{FC}$ grows due to large distance (larger than Mpc) or equivalency  large time scale  of evolution (see Eq. \ref{fc}).
In principle, under effects of background fields, particle
scattering and temperature fluctuations, linear polarizations
of CMB radiation field propagating from the last scattering
surface can rotate each other and convert to circular
polarizations. In this study we will study the distribution of neutrino-photon scattering for the generation of the circular polarization of the CMB. First we give a brief introduction on Stokes parameters and derive the time evolution of these parameters in terms of the photon-particle scattering. Then by considering the weak  and electrodynamic interactions, we will find the time evolution of Stokes parameters in terms of  photon-neutrino interactions. Finally we try to estimate the maximum value of the $V$-mode polarization by using the relevant values of energy and number density of cosmic neutrinos around recombination epoch.

\section{Stokes parameters}
As usual, we characterized the polarization of CMB by means of the Stokes parameters
of radiation: I, Q, U and V.
Assume a  quasi-monochromatic electromagnetic wave propagating in
the $\hat z$-direction which 
is described by:
 \bea
E_x=a_x(t)\cos[\omega_0t-\theta_x(t)],\quad
E_y=a_y(t)\cos[\omega_0t-\theta_y(t)],\eea
where amplitudes $a_{x,y}$ and phase angles $\theta_{x,y}$ are slowly varying functions with respect to the period ${\mathcal T}_0=2\pi/\omega_0$.
Stokes parameters, which describe polarization states of a nearly monochromatic electromagnetic wave,
are defined as the following time averages \cite{jackson}:
\bea I &=&\langle a^2_
x \rangle+ \langle a^2_y\rangle, \nonumber\\ Q &=&\langle a^2_ x
\rangle- \langle a^2_y\rangle, \nonumber\\ U &=&\langle 2a_ x
a_y\cos(\theta_x-\theta_y)\rangle,\nonumber\\ V &=&\langle 2a_ x
a_y\sin(\theta_x-\theta_y)\rangle,
\label{ic}
\eea
where the parameter $I$
is total intensity, $Q$ and $U$ intensities of linear
polarizations of electromagnetic  waves, whereas the $V$
parameter indicates
the difference between left- and right- circular polarizations
intensities. Linear polarization can also be characterized through
a vector of modulus $P_L\equiv\sqrt{Q^2+U^2}$. The time evolution of these Stokes parameters
is given through Boltzmann equation. Boltzmann equation is a systematic
mechanism in order to describe the evolution of the distribution function under gravity
and collisions. Ones can consider each polarization state of the CMB radiation as the phase space
distribution function $\xi$. The classical Boltzmann equation generally is written as
\begin{equation}\label{B-E}
    \frac{d}{dt}\xi=\mathcal{C}(\xi)
\end{equation}
where the left hand side is known as the Liouville term deals with the effects of gravitational perturbations about the homogeneous cosmology. The right hand side of the Boltzmann
equation contains all possible collision terms. By considering the contribution of the neutrino-photon scattering on the right hand side of above equation, we calculate the time evolution of  the each polarization state of the CMB photons.
For
the rest of calculations, Stokes parameters are given in a quantum-mechanical description. An arbitrary polarized
state of a photon $(|k^0|^2=|{\bf k}|^2)$, propagating in the
$\hat z$-direction, is given by \bea
|\epsilon\rangle=a_1\exp(i\theta_1)|\epsilon_1\rangle+a_2\exp(i\theta_2)|\epsilon_2\rangle,\eea
where linear bases $|\epsilon_1\rangle$ and
$|\epsilon_2\rangle$ indicate the polarization states in the $x$-
and $y$-directions. Quantum-mechanical operators in this
linear bases, corresponding to Stokes parameter, are given by
\bea
\hat{I}&=&|\epsilon_1\rangle\langle\epsilon_1|+|\epsilon_2\rangle\langle\epsilon_2|,\nonumber\\
\hat{Q}&=&|\epsilon_1\rangle\langle\epsilon_1|-|\epsilon_2\rangle\langle\epsilon_2|,\nonumber\\
\hat{U}&=&|\epsilon_1\rangle\langle\epsilon_2|+|\epsilon_2\rangle\langle\epsilon_1|,\nonumber\\
\hat{V}&=&i|\epsilon_2\rangle\langle\epsilon_1|-i|\epsilon_1\rangle\langle\epsilon_2|.
\label{i-v} \eea An ensemble of photons in a general mixed state
is described by a normalized density matrix $\rho_{ij}\equiv
(\,|\epsilon_i\rangle\langle \epsilon_j|/{\rm tr}\rho)$, and the dimensionless
expectation values for Stokes parameters are given by
\bea
I\equiv\langle  \hat I \rangle &=& {\rm tr}\rho\hat I
=1,\label{i}\\
Q\equiv\langle  \hat Q \rangle &=& {\rm tr}\rho\hat{Q}=\rho_{11}-\rho_{22},\label{q}\\
U\equiv\langle
 \hat U\rangle &=&{\rm tr}\rho\hat{U}=\rho_{12}+\rho_{21},\label{u}\\
V\equiv\langle  \hat V \rangle &=& {\rm
tr}\rho\hat{V}=i\rho_{21}-i\rho_{21}, \label{v}
\eea
where ``$\rm tr$'' indicates the trace in the space of polarization states. These above equations determine
the relationship between four Stokes parameters and the
$2\times 2$ density matrix $\rho$ of photon polarization states. In this section, we use  notations which used in \cite{xue}.


\section{The generation of polarized CMB via photon-neutrinos scattering.}
\begin{figure}
\begin{center}
  \includegraphics[width=0.6\columnwidth]{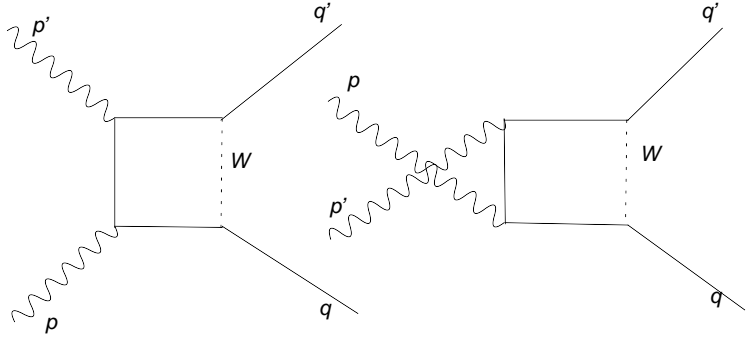}\\
  \caption{The typical diagrams of photon-neutrino scattering is given in this plot. }\label{cmb1}
  \end{center}
\end{figure}

The density operators describing a system of photons is given by
\bea
\hat\rho=\frac{1}{\rm {tr}(\hat \rho)}\int\frac{d^3k}{(2\pi)^3}
\rho_{ij}(k)a^\dagger_i(k)a_j(k),
\eea
where $\rho_{ij}(k)$ is the general density-matrix (\ref{i}-\ref{v}) in the space of polarization states with a fixed energy-momentum ``$k$''. The number operator $
D^0_{ij}(k)\equiv a_i^\dag (k)a_j(k)$.
 Then the
expectation value of this number operator is defined by
\bea
\langle\, D^0_{ij}(k)\,\rangle\equiv {\rm tr}[\hat\rho
D^0_{ij}(k)]=(2\pi)^3 \delta^3(0)(2k^0)\rho_{ij}(k).\label{t1}
\eea
And on the other hand, the time evolution of the operator $D^0_{ij}(k)$, considered in the Heisenberg picture, is
\begin{equation}\label{heisen}
   \frac{d}{dt} D^0_{ij}(k)= i[H,D^0_{ij}(k)],
\end{equation}
where $H$ is the full Hamiltonian. Taking the expectation value of both sides of  above equation gives the Boltzmann equation (\ref{B-E}) for the system's density matrix (as well as polarization states) which is a generalization of the usual classical Boltzmann equation for particle occupation numbers. By substituting Eq.~(\ref{t1}) in Eq.~(\ref{heisen}), the time evolution of $\rho_{ij}(k)$ as well as Stokes parameters is given \cite{cosowsky1994},
\bea
(2\pi)^3 \delta^3(0)(2k^0)
\frac{d}{dt}\rho_{ij}(k) = i\langle \left[H^0_I
(t);D^0_{ij}(k)\right]\rangle-\frac{1}{2}\int dt\langle
\left[H^0_I(t);\left[H^0_I
(0);D^0_{ij}(k)\right]\right]\label{bo}\rangle,
\eea
where $H^0_I(t)$ is the first order of the interacting Hamiltonian.  The first term on the right-handed side is a forward scattering term, and the second one is a higher order collision term.
In order to find effects of photon- neutrinos scattering on the polarization of the CMB, we  start with following lagrangian
\bea\label{lagw}
\pounds_{I} = \pounds_{_{QED}} +
\pounds_{_{e\nu}},
\eea
where the first term $\pounds_{_{QED}}$  is the quantum
electrodynamic lagrangian (QED),
and the second term $\pounds_{_{e\nu}}$ is the lagrangian of weak interaction containing electron-neutrino vertex. In context of standard model, there is no direct vertex for photon-neutrino however the first order of the interaction between photon-neutrino appears during one-loop interaction where photons and neutrinos both interact with electrons and weak gauge bosons (see Fig.~\ref{cmb1}).

We express the electromagnetic free gauge field $A_\mu$ in terms of plane wave solutions in the Coulomb gauge \cite{zuber}, 
 \beq A_\mu(x) = \int \frac{d^3 k}{(2\pi)^3
2 k^0} \left[ a_i(k) \epsilon _{i\mu}(k)
        e^{-ik\cdot x}+ a_i^\dagger (k) \epsilon^* _{i\mu}(k)e^{ik\cdot x}
        \right],\eeq
where $\epsilon _{i\mu}(k)$ are the polarization
four-vectors and the index $i=1,2$, representing two transverse polarizations of a free photon with four-momentum $k$ and $k^0=|{\bf{k}}|$. $a_i(k)$
$[a_i^\dagger (k)]$ are the creation [annihilation] operators, which satisfy the canonical commutation relation as following
\begin{equation}
        \left[  a_i (k), a_j^\dagger (k')\right] = (2\pi )^3 2k^0\delta_{ij}\delta^{(3)}({\bf k} - {\bf k}' ).
\label{comm}
\end{equation}
Also the free fermion field $\psi$ is given:
 \beq
 \psi(x) = \int \frac{d^3 p}{(2\pi)^3}\frac{1}{
\sqrt{2 E_{\mathbf{p}}}} \sum_r\left[ b_r(p) U_{r}(p)
        e^{-ip\cdot x}+ d_r^\dagger (p) \mathcal{V}_{r}(p)e^{ip\cdot x}
        \right],\label{psi}
\eeq
where $U_r$ and $\mathcal{V}_{r}$ are Dirac spinors,  $b_r$ ($d_r$) and  $b^\dagger_r$ ($d^\dagger_r$) are creation and inhalation operators for fermions (anti-fermions), which satisfy following relations,
\begin{equation}
        \left\{  b_s (p), b_r^\dagger (p')\right\}=\left\{  d_s (p), d_r^\dagger (p')\right\} = (2\pi )^3 \delta_{sr}\delta^{(3)}({\bf p} - {\bf p}' ).
\label{comm}
\end{equation}
\comment{The density operator describing an ensemble of free photons in the space of energy-momentum and polarization state is given by 
\bea
\langle\, D^0_{ij}(k)\,\rangle\equiv {\rm tr}[\hat\rho
D^0_{ij}(k)]=(2\pi)^3 \delta^3(0)(2k^0)\rho_{ij}(k).
\eea}
Now by using Lagrangian (\ref{lagw}) and Fig.1, first order photon-neutrino Hamiltonian interaction  is given by
\begin{eqnarray}
  H^0_I &=& \int d\mathbf{q} d\mathbf{q'} d\mathbf{p} d\mathbf{p'} (2\pi)^3\delta^3(\mathbf{q'} +\mathbf{p'} -\mathbf{p} -\mathbf{q} ) \nonumber \\
   &\times& \exp[it(q'^0+p'^0-q^0-p^0)]\left(b^\dagger_{r'}a^{\dagger}_{s'}(\mathcal{M}_1+\mathcal{M}_2)a_sb_r\right)
\end{eqnarray}

where
\begin{eqnarray}
  \mathcal{M}_1+\mathcal{M}_2 &=& -\frac{1}{8}e^2g_w^2\int\frac{d^4k}{(2\pi)^4}D_{\alpha\beta}(q-k)\bar{U}_{r'}(q')\gamma^\alpha (1-\gamma_5)S_F(k+p-p')\nonumber\\
   &\times& \left[\ep_{s'}S_F(k+p)\ep_{s}+\ep_{s}S_F(k-p')\ep_{s'}\right]
   S_F(k)\gamma^\beta (1-\gamma_5)U_r(q),
\end{eqnarray}
here $D_{\alpha\beta}$ and $S_F$ are boson and fermion propagators, $g_W$ is the weak coupling constant and our notation
$d\mathbf{q}=d^3q/[(2\pi)^32q^0]$, the same for $d\mathbf{p},d\mathbf{p'}$ and $d\mathbf{q'}$. By using above result $H^0_I$ and Eq.(\ref{bo}), we are ready to find the commutator in the photon-neutrino forward scattering term
\begin{eqnarray}
  [H^0_I, D^0_{ij}({\bf k})] &=& \int d\mathbf{q} d\mathbf{q'} d\mathbf{p} d\mathbf{p'} (2\pi)^3\delta^3(\mathbf{q'} +\mathbf{p'} -\mathbf{p} -\mathbf{q} ) (\mathcal{M}_1+\mathcal{M}_2)\nonumber\\
  &\times& (2\pi)^3[b^\dagger_{r'}b_{r}a^\dagger_{s'}a_{s}2p^0\delta_{is}\delta^3({\bf k}-{\bf p})-b^\dagger_{r'}b_{r}a^\dagger_{i}a_{s}2p'^0\delta_{js'}\delta^3({\bf k}-{\bf p'})].\label{fws}
\end{eqnarray}
On using the above expectation values and below operator expectation values \cite{cosowsky1994},
\bea \langle \, a_1a2...b_1b_2...\, \rangle
&=&\langle \,a_1a_2...\, \rangle\langle \,b_1b_2...\, \rangle
\label{contraction1}\\
\langle \, a^\dag_{s'}(p')a_{s}(p)\, \rangle
&=&2p^0(2\pi)^3\delta^3(\mathbf{p}-\mathbf{p'})\rho_{ss'}(\mathbf{x},\mathbf{p}),
\label{contraction2}\\
\langle \, b^\dag_{r'}(q')b_{r}(q)\, \rangle
&=&(2\pi)^3\delta^3(\mathbf{q}-\mathbf{q'})\delta_{ss'}\frac{1}{2}n_\nu(\mathbf{x},\mathbf{q}),
\label{contraction3} \eea
it follows that
\begin{eqnarray}
  i\langle[H^0_I, D^0_{ij}({\bf k})]\rangle &=& -\frac{i}{16}e^2g_w^2\int d\mathbf{q} (\rho_{s'j}({\bf k})\delta_{is} -\rho_{is}({\bf k})\delta_{js'})n_\nu(\mathbf{x},\mathbf{q})\nonumber\\
  &\times& \int\frac{d^4l}{(2\pi)^4}D_{\alpha\beta}(q-l)\bar{U}_{r}(q')\gamma^\alpha (1-\gamma_5)S_F(l)\nonumber\\
   &\times& \left[\,\ep_{s'},S_F(l+k)\,\ep_{s}+\ep_{s}\,S_F(l-k)\ep_{s'}\,\,\right]
   S_F(l)\gamma^\beta (1-\gamma_5)U_r(q),\label{fws1}
\end{eqnarray}
where integrating on $l$ comes from the loop interaction of photon-neutrino and $n_\nu(\mathbf{x},\mathbf{q})$ represents the number density of neutrinos of momentum $\mathbf{q}$ per unit volume (C$\nu$B distribution function).
\comment{ so that
\begin{equation}\label{nd-n0}
    n_\nu(\mathbf{x})=\frac{1}{(2\pi)^3}\int d^3q \,\,n_\nu(\mathbf{x},\mathbf{q}),
\end{equation}
where $n_\nu(\mathbf{x})$ is the local neutrino number density. And also the average bulk momentum of neutrinos in $\hat{i}$-direction is given by
\begin{equation}\label{nd-n1}
   <q_i>=\frac{1}{ n_\nu(\mathbf{x})}\int\frac{ d^3q }{(2\pi)^3}\,n_\nu(\mathbf{x},\mathbf{q})\,q_i=\tilde{q}_i=m_\nu \tilde{v}_i,\,\,\,\,\, i=x,y,z,
\end{equation}
where $\mathbf{\tilde{q}}\neq0$ is the average bulk momentum  of neutrino and $|\mathbf{\tilde{v}}|$ is the neutrino bulk velocity which we consider $|\mathbf{\tilde{v}}|< \Delta T/T$.}
 By helping Dimensional regularization and Feynman parameters, we go forward to obtain  the leading order term of the right side of the above equation, then
\begin{eqnarray}
  i\langle[H^0_I, D^0_{ij}({\bf k})]\rangle &=& -\frac{1}{16}\frac{1}{4\pi^2}e^2g_w^2\int d\mathbf{q} (\rho_{s'j}({\bf k})\delta_{is} -\rho_{is}({\bf k})\delta_{js'})n_\nu(x,q)\nonumber\\
  &\times& \int_0^1dy\int_0^{1-y}dz\frac{(1-y-z)}{zM^2_W}\bar{U}_{r}(q)(1+\gamma_5)(2z q\!\!\!/\epsilon_{s'}.\epsilon_s\nonumber\\
   &+& \,2z(\ep_{s'}\,\,\mathbf{q}.\epsilon_s\,+\ep_{s}\,\mathbf{q}.\epsilon_{s'}\,)+
   (3y-1)k\!\!\!/\,(\ep_{s}\,\ep_{s'}\,-\ep_{s'}\,\ep_{s}\,)\,)
  U_r(q).\label{fws2}
\end{eqnarray}
Here we use the gamma-matrix identity $A\!\!\!\!/\,B\!\!\!\!/=2A.B-B\!\!\!\!/\,A\!\!\!\!/$, the polarization vector properties $k.\epsilon_i=0$ and
$\epsilon_i.\epsilon_j=-\delta_{ij}$. Now every thing is ready to see the time evolution of stocks parameters as well as each polarization state of CMB photons. We are interested in $V$ parameter which gives the contribution of the circular polarization, by considering Eqs.(\ref{bo},\ref{fws2}),  $dV/dt$ is given as following
\begin{eqnarray}
  \frac{dV(\mathbf{x},\mathbf{k})}{dt} &\approx& \frac{1}{6}\frac{1}{(4\pi)^2}\frac{e^2g_w^2}{M^2_W k^0}\int d\mathbf{q}\,\,n_\nu(x,q)\bar{U}_{r}(q)(1+\gamma_5) \nonumber\\
  &\times& \left[(\ep_{1}\,q.\epsilon_1-\ep_{2}\,q.\epsilon_2)Q(\mathbf{k})-
  (\ep_{1}\,q.\epsilon_2+\ep_{2}\,q.\epsilon_1)U(\mathbf{k})\right]U_r(q),\label{v1}
\end{eqnarray}
we should remind that $U(k)$ is one of stocks parameters which represents linear polarization while $U_r$ is Dirac spinor.  We neglect the terms with  $1/M_w^4$ order and smaller than one.
In order to go further, let's introduce Dirac spinors and our frame work in more details. $\gamma^\mu$, $\gamma^5$ and $U_r(q)$ are given

\begin{equation}\label{gamma}
   \gamma^\mu=\left(
  \begin{array}{cc}
    0 & \sigma^\mu \\
    \bar{\sigma}^\mu & 0 \\
  \end{array}
\right),\,\,\,\,\,\,\gamma^5=\left(
  \begin{array}{cc}
    -I & 0 \\
    \,0& I \\
  \end{array}
\right),\,\,\,\,\,
U_r(q)=\left(
         \begin{array}{c}
           \sqrt{p.\sigma} \xi^r\\
           \sqrt{p.\bar{\sigma}} \xi^r\\
         \end{array}
       \right),
\end{equation}
where $\sigma^\mu=(1,\vec{\sigma})$, $\bar{\sigma}^\mu=(1,-\vec{\sigma})$ and $\xi$ is two-component spinor normalized to unity and  also  belove equations are useful
\begin{equation}\label{sn2}
   \bar{U}_r(q)\gamma^\mu U_s(q)=2q^\mu\delta^{rs},\,\,\,\,\, \frac{1}{2}\sum_r\bar{U}_r(q)\gamma^\mu(1-\gamma^5) U_r(q)=2q^\mu.
\end{equation}
Then by making average on the spin of neutrinos, $\frac{1}{2}\sum_r$, and substituting above equations in Eq.(\ref{v1}), we arrive
\begin{eqnarray}
  \frac{dV(\mathbf{x},\mathbf{k})}{dt} &\approx& \frac{\sqrt{2}}{3\pi k^0}\alpha\,G^F\int d\mathbf{q}\,\,n_\nu(x,q) \nonumber\\
  &\times& \left[\,(q.\epsilon_{1}\,\,q.\epsilon_1-q.\epsilon_{2}\,\,q.\epsilon_2)Q(\mathbf{k})-
  (q.\epsilon_{1}\,\,q.\epsilon_2+q.\epsilon_{2}\,\,q.\epsilon_1)U(\mathbf{k})\,\right],\label{v2}
\end{eqnarray}
where
\begin{equation}\label{cof0}
   G^F=\frac{\sqrt{2}}{8}\frac{g_W^2}{M_W^2}\approx1.16\times10^{-5}(GeV)^{-2},\,\,\,\,\alpha=\frac{e^2}{4\pi}=1/137.
\end{equation}
This equation contains an integration on neutrinos momentum which should determine. In next section, by introducing C$\nu$B distribution function and perturbations, we try to estimate the integration in (\ref{v2}).

\section{ C$\nu$B distribution function and perturbations}
\begin{figure}
  \includegraphics[width=4in]{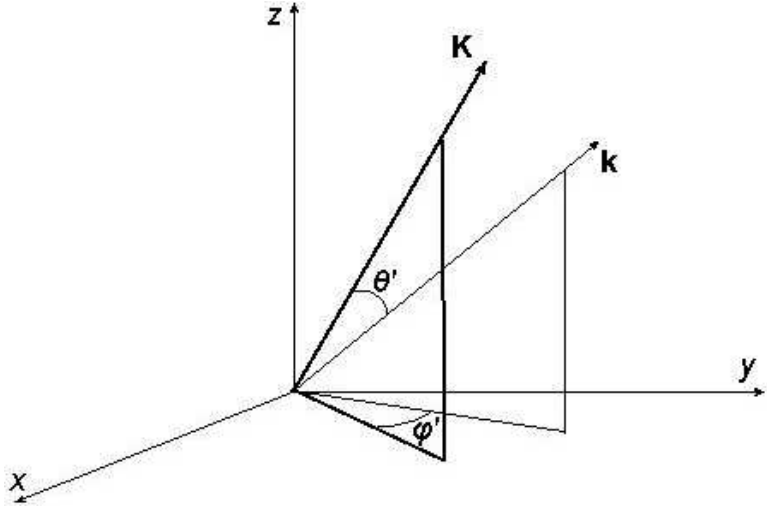}\\
  \caption{Angles and directions to determine the angular dependence of equation (\ref{v7'}) is given in above plot.}\label{cmb3}
\end{figure}

It is convenient to write the phase space distribution of neutrino $f_{\nu}(\vec{\mathbf{x}},\vec{\mathbf{q}},\tau)$ as a
zeroth-order distribution $f_{\nu0}(\vec{\mathbf{x}},\vec{\mathbf{q}},\tau)$ plus a perturbed piece $\Psi(\vec{\mathbf{x}},\vec{\mathbf{q}},\tau)$ as following \cite{MC,Ma,lesg}
\begin{equation}\label{nu-DF}
    f_{\nu}(\vec{\mathbf{x}},\vec{\mathbf{q}},\tau)=f_{\nu0}[1+\Psi(\vec{\mathbf{x}},\vec{\mathbf{q}},\tau)],
\end{equation}
where $\vec{\mathbf{q}}=q \hat{\mathbf{n}}$ and $\hat{\mathbf{n}}$ indicates the direction of neutrino velocity.
This phase space distribution evolves according to the Boltzmann equation. In terms of our
variables $(\vec{\mathbf{x}},q,\hat{\mathbf{n}},\tau)$, we have
\begin{equation}\label{bonu}
    \frac{\partial f_\nu}{\partial \tau}+i \frac{q}{ \varepsilon_\nu}(\vec{\mathbf{K}}\cdot\hat{\mathbf{n}})\Psi+
     \frac{d\ln f_{\nu0}}{d\ln q}[\dot\phi-i \frac{ \varepsilon_\nu}{q}(\vec{\mathbf{K}}\cdot\hat{\mathbf{n}})\psi]=0
\end{equation}
where $\mathbf{K}$ (wave number) is the Fourier conjugate of $\mathbf{x}$, the collision terms on the RHS of Eq.(\ref{bonu}) is neglected due to
weak interactions of neutrino and
two scalar potentials $\phi$ and $\psi$ characterize metric perturbations, they appear
in the line element as
\begin{equation}\label{line}
    ds^2=a^2(\tau)\{-(1+2\psi)d\tau^2+(1+2\phi)dx_i\,dx^j\}.
\end{equation}
Notice, the terms in the Boltzmann equation depend on the direction of the momentum $\hat{n}$ only
through its angle with $\vec{\mathbf{K}}$ ($\mu'=\hat{\mathbf{K}}\cdot\hat{n}$).
The conformal Newtonian gauge (also known as the longitudinal gauge) advocated by
Mukhanov et al.\cite{mukhanov} is a particularly simple gauge to use for the scalar mode of metric
perturbations. It should be emphasized that the conformal Newtonian gauge is a restricted gauge since the metric
is applicable only for the scalar mode of the metric perturbations; the vector and the tensor
degrees of freedom are eliminated from the beginning. By considering the collision-less Boltzmann equation (\ref{bonu}) and expanding the angular
dependence of the perturbation $\Psi$ in a series of Legendre polynomials $P_l(\mu')$ as following
\begin{equation}\label{App1}
  \Psi(\vec{\mathbf{K}},q,\mu',\tau)=\sum_{l=0} (-i)^l(2l+1)\Psi_{ l}(\vec{\mathbf{K}},\tau)P_l(\mu'),
\end{equation}
Instead of assuming a form for $f_\nu$, it will be helpful taking the perturbations of energy density $\delta\rho_{\nu}$, pressure $\delta P_{\nu}$, energy flux $\delta T^0_{\nu i}$, and shear stress  as following \cite{Ma}
\begin{eqnarray}
  \delta\rho_{\nu} &=&4\pi a^{-4}\int\, q^2\, dq\,\varepsilon_\nu\,f_{\nu0}\Psi_0,\,\,\,\,\,\,\,\,\,\,\, \delta P_{\nu} =\frac{4\pi}{3} \,a^{-4}\int\, q^2\, dq\,\frac{q^2}{\varepsilon_\nu}f_{\nu0}\Psi_0 \nonumber \\
 (\bar\rho_{\nu}+\bar P_{\nu})\theta_{\nu} &=&4\pi a^{-4}\int\, q^2\, dq\,\,q\,f_{\nu0}\Psi_1,\,\,\,\,\,\, (\bar\rho_{\nu}+\bar P_{\nu})\sigma_{\nu} =\frac{8\pi}{3} \,a^{-4}\int\, q^2\, dq\,\frac{q^2}{\varepsilon_\nu}\,f_{\nu0}\Psi_2,
\end{eqnarray}
where $\varepsilon_\nu=(q^2+a^2\,m_\nu^2)^{1/2}$ ($m_\nu\simeq 0.1$eV) and $a$ is scalar factor. Next we can rewrite Eq.(\ref{v2}) in terms of $\Psi_{ l}$ and $\mu$ (indicates the angle between CMB photons direction and $\vec{\mathbf{K}}$) as following
\begin{eqnarray}
 &&\frac{1}{(2\pi)^3}\left(\,\left(\hat{\mathbf{K}}\cdot(\mathbf{\epsilon}_{1}-\mathbf{\epsilon}_{2})\right)^2\,Q-
  2\hat{\mathbf{K}}\cdot\mathbf{\epsilon}_{1}\,\,\hat{\mathbf{K}}\cdot\mathbf{\epsilon}_2\,U\right)\int q^2\, dq\,d\Omega\, \mu'^2\, \frac{q^2}{\varepsilon_\nu}\, f_{\nu0}\Psi(\vec{\mathbf{K}},q,\mu',\tau),\nonumber\\
  &=&\frac{1}{(2\pi)^3}\left(\,\left(\hat{\mathbf{K}}\cdot(\mathbf{\epsilon}_{1}-\mathbf{\epsilon}_{2})\right)^2\,Q-
  2\hat{\mathbf{K}}\cdot\mathbf{\epsilon}_{1}\,\,\hat{\mathbf{K}}\cdot\mathbf{\epsilon}_2\,U\right)\frac{4\pi}{3}\int q^2\, dq\,\frac{q^2}{\varepsilon_\nu}\, f_{\nu0}[\Psi_0-2\Psi_2],\nonumber\\
  &\simeq&\left(\,\left(\hat{\mathbf{K}}\cdot(\mathbf{\epsilon}_{1}-\mathbf{\epsilon}_{2})\right)^2\,Q-
  2\hat{\mathbf{K}}\cdot\mathbf{\epsilon}_{1}\,\,\hat{\mathbf{K}}\cdot\mathbf{\epsilon}_2\,U\right)\tilde\rho_{\nu}
  \left[\frac{\delta P_{\nu}}{\tilde\rho_{\nu}} -2\sigma_{\nu}\right]
  \label{int00}
\end{eqnarray}
where
\begin{eqnarray}
  \dot{\Psi}_0 &=& -\frac{qK}{\varepsilon_\nu}\Psi_1-\dot{\phi}\frac{d\ln f_{\nu0}}{d\ln q} \nonumber \\
  \dot{\Psi}_1&=& \frac{qK}{3\varepsilon_\nu}(\Psi_0-2\Psi_2)+\frac{\varepsilon_\nu\,K}{3q}\psi \frac{d\ln f_{\nu0}}{d\ln q}\nonumber \\
  \dot{\Psi}_l &=& \frac{qK}{(2l+1)\varepsilon_\nu}(l\Psi_{l-1}-(l+1)\Psi_{l+1}),\,\,\,\,\,\,l\geq2,
\end{eqnarray}
and
\begin{equation}\label{rohbar}
  \tilde\rho_{\nu} =\frac{1}{2\pi^2}\, a^{-4}\int\, q^2\, dq\,\varepsilon_\nu\,f_{\nu0},\,\,\,\,\,\,\,\,\,\tilde{P}_{\nu} =\frac{1}{6\pi^2}\, a^{-4}\int\, q^2\, dq\,\frac{q^2}{\varepsilon_\nu}\,f_{\nu0},
\end{equation}
The evolution equations derived for neutrino perturbations can be solved numerically once the
initial perturbations are specified. By starting the integration at early times when a given K-mode
is still outside the horizon $K\tau<<1$ and implementing very basic iso-curvature and adiabatic initial conditions given by \cite{Ma},
we can obtain an estimation for the value of the perturbations. The value of $\eta_\nu=\frac{\delta P_{\nu}}{\tilde\rho_{\nu}} -2\sigma_{\nu}$ depends on  time or red-shift and the wave number $K$, but for simplicity we consider the time average value of this quantity at $K=K_*=0.05/Mpc$ which is given by
\begin{equation}\label{pertur}
  \bar\eta_\nu=<[\frac{\delta P_{\nu}}{\tilde\rho_{\nu}} -2\sigma_{\nu}]>=\frac{1}{\tau_0-\tau_{lss}}\int_{\tau_0}^{\tau_{lss}}\,d\tau\,[\frac{\delta P_{\nu}}{\tilde\rho_{\nu}} -2\sigma_{\nu}(\tau)](\tau,K_*)\simeq 0.3\frac{\Delta T}{T}\big|_{\nu}\leq 10^{-5},
\end{equation}
where $\tau_{lss}$ indicates the time at last scattering surface and $\tau_0$ is present time.
Next by using Eqs.(\ref{int00}), (\ref{pertur}), the time evolution of $V$ mode is given
\begin{eqnarray}
  \frac{dV(\mathbf{x},\mathbf{k})}{dt} &\approx& \frac{\sqrt{2}}{3\pi}\,\frac{1}{k^0}\alpha\,G^F\, \tilde{\rho}_\nu(\mathbf{x})\bar\eta_\nu\nonumber\\
  &\times&\left(\,\left(\hat{\mathbf{K}}\cdot(\mathbf{\epsilon}_{1}-\mathbf{\epsilon}_{2})\right)^2\,Q-
  2\hat{\mathbf{K}}\cdot\mathbf{\epsilon}_{1}\,\,\hat{\mathbf{K}}\cdot\mathbf{\epsilon}_2\,U\right).\label{v3}
\end{eqnarray}
We go forward by considering $k^0\approx T_\gamma$ and also to avoid from the angular distribution of each mode
\begin{equation}\label{v4}
    V(\mathbf{x},k)=\int\frac{ d\Omega}{4\pi}V(\mathbf{x},\mathbf{k}),\,\,\,\,\,k=|\mathbf{k}|=k^0,
\end{equation}
where  $d\Omega$ is  the differential solid angle. We consider  $Q(\mathbf{x},k)$, $U(\mathbf{x},k)$ and $I(\mathbf{x},k)$ modes in the same way as well as $V(\mathbf{x},k)$. Then
\begin{eqnarray}
  \frac{dV}{dt}(\mathbf{x},k) &\approx& \frac{\sqrt{2}}{3\pi}\,\frac{1}{k^0}\alpha\,G^F\, \tilde{\rho}_\nu(\mathbf{x})\bar\eta_\nu\, (C_U+C_Q)\label{v5}
\end{eqnarray}
where
 \begin{eqnarray}
 C_U&= &-2\int\frac{ d\Omega}{4\pi}(\mathbf{\hat{K}}.\hat{\epsilon}_{1}\,\,\mathbf{\hat{K}}.\hat{\epsilon}_2)\,\label{cu}\\
C_Q&= &\int\frac{ d\Omega}{4\pi} \,(\mathbf{\hat{K}}.\hat{\epsilon}_{1}\,\,\mathbf{\hat{K}}.\hat{\epsilon}_1-
  \mathbf{\hat{K}}.\hat{\epsilon}_{2}\,\,\mathbf{\hat{K}}.\hat{\epsilon}_2)Q(\mathbf{x},\mathbf{k}).\label{cq}
\end{eqnarray}
To estimate the $V$ mode, we integrate over time $\int dt=\int dz\,a/H(z)$, where the redshift $z\in [0, 10^3]$, the Hubble function $H(z)=H_0[\Omega_M(z+1)^3+\Omega_\Lambda)]^{1/2}$ for $\Omega_M\simeq 0.3$, $\Omega_\Lambda\simeq 0.7$ and $H_0=75$ km/s/Mpc, and the temperature
$T_{\gamma,\nu} = T_{0,\gamma,\nu}(1+z)$ [$T_{0,\gamma}\approx 2.725 K^\circ=2.349\times 10^{-4}{\rm
eV}=(0.511{\rm cm})^{-1}$] in the standard cosmology \cite{MC}. And also $\tilde{\rho}_\nu=\tilde{\rho}_\nu^0\,F(z)$ where $\tilde{\rho}_\nu^0\simeq m_\nu n_\nu^0(\mathbf{x})\sim  m_\nu 112/cm^3$ is today's C$\nu$B density and
\begin{equation}\label{fz}
  F(z)=\frac{4\pi}{\tilde{\rho}_\nu^0}\,(1+z)^{4}\,\int q^2\,dq\,\left(q^2+(1+z)^{-2}\,m_\nu^2\right)^{1/2}\, f_{\nu0}.
\end{equation}
Finally $V$ is given by
\begin{eqnarray}
  V&\approx& \frac{\sqrt{2}}{3\pi}\,\frac{1}{k^0}\alpha\,G^F\, \tilde{\rho}^0_\nu\,\bar\eta_\nu\, \int_0^{1000}\frac{dz}{(1+z)^2}\frac{F(z)}{H(z)}  (C_U+C_Q),\label{v6}
\end{eqnarray}
where we can substitute
\begin{equation}\label{cofi}
   H_0^{-1}\approx6\times10^{41}(GeV)^{-1},\,\,\,\,\,\, T_{0,\nu}\approx1.67\times10^{-13}GeV,\,\,\,\,T_{0,\gamma}\approx2.34\times10^{-13}GeV,
\end{equation}
into equation (\ref{v6}) and integrate on redshift $z$. By assuming the independence $(C_U+C_Q)$ from redshift, we arrive
\begin{eqnarray}
  V (\mathbf{x},k)&\approx& \frac{1}{2}\left(\frac{T_{0,\gamma}}{k}\right)\left(\frac{m_\nu}{0.1eV}\right)\left(\frac{n^0_\nu(\mathbf{x})}{112/cm^3}\right)\,\bar\eta_\nu\,(C_U+C_Q),\label{v7}
\end{eqnarray}
where $k\simeq T_{0,\gamma}$ is the values of average energy of CMB at present universe.
Now we need to have some estimations of $C_U$  and $C_Q$ as well as $U$ and $Q$ polarization modes. First approximation for these quantities is given by knowing  that $V/I<\delta T/T$  \cite{anisotropy}, this implies that $\bar\eta_\nu\,(C_U+C_Q)/I$ should be smaller than $10^{-7}$. Let's investigate $C_U$ and $C_Q$ more precisely. A Fourier transform over the spatial dependence $\mathbf{x}$ of the equation (\ref{v7}) gives
\begin{eqnarray}
  V (\mathbf{K},k)&\approx& 10^{2}\left(\frac{T_{0,\gamma}}{k}\right)\left(\frac{m_\nu}{0.1eV}\right)\left(\frac{n^0_\nu}{112/cm^3}\right)\,\bar\eta_\nu\,\left(C_U(\mathbf{K},k)+C_Q(\mathbf{K},k)\right).\label{v7'}
\end{eqnarray}
 For the rest of paper by neglecting the effects of tensor perturbations and assuming  $\theta'$ as the angel between the $\mathbf{K}$ mode perturbation  and the photon direction $\mathbf{k}$ [see Fig.(\ref{cmb3})], then we have
\begin{eqnarray}
 C_U(\mathbf{K},k)&= &-\int\frac{ d\Omega'}{4\pi} (\sin^2\theta'\,\sin2\phi')U(\mathbf{K},\mathbf{k})\,\label{cu1}\\
C_Q(\mathbf{K},k)&= &\int\frac{ d\Omega'}{4\pi}(\sin^2\theta'\,\cos2\phi')Q(\mathbf{K},\mathbf{k}),\label{cq1}
\end{eqnarray}
where $k=|\mathbf{k}|$. Then ones can expand the incident intensities $U$ and $Q$ in spherical
harmonics (around  $\mathbf{K}$ direction) as following
\begin{equation}\label{har}
    U(\mathbf{K},\mathbf{k})=\sum_{lm}\mathfrak{u}_{lm}(k)Y_{l,m}(\theta',\,\phi'),\,\,\,\,Q(\mathbf{K},\mathbf{k})=\sum_{lm}\mathfrak{q}_{lm}(k)Y_{l,m}(\theta',\,\phi').
\end{equation}
By using equations (\ref{v7})-(\ref{har}), the photon-neutrino scattering generates circular polarization CMB from initially
linear polarized CMB if this linear intensity ($C_U$ and $C_Q$) at a given point as a function of direction has no-zero component $Y_{22}$.
\begin{eqnarray}
 C_U(\mathbf{K},k)&= &-\frac{2}{\pi} (\sqrt{\frac{2\pi}{15}})\mathfrak{u}_{22}(k)\,\label{cu2}\\
C_Q(\mathbf{K},k)&= &\frac{2}{\pi} (\sqrt{\frac{2\pi}{15}})\mathfrak{q}_{22}(k).\label{cq2}
\end{eqnarray}
As a result of this calculation, the exactly value of $V$-parameter in each $\mathbf{K}$-mode  depends to the quadruple components of the incident intensity distribution and  the C$\nu$B perturbations $\delta P_{\nu}$ and $\sigma_\nu$.

\section{ The anisotropy of the photon distribution.}

The pervious section has devoted for the right side of the Boltzman equation, collision and scattering terms. In this section we discuses the lift side term which describes the propagation of photons in the background space-time. As discussed in \cite{cosowsky1994,zal,hu,MC}, the first order deviation from flat space-time in the metric perturbation leads to an incommodiously  for the photon CMB distribution function which is necessary to generate unpolarized CMB due to Thomson scattering.
To go further, we expand the photon distribution function $f_\gamma(\mathbf{k},\mathbf{x})$ about its
zero-order Bose-Einstein value as following
\begin{eqnarray}
 f_\gamma(\mathbf{k},\mathbf{x},t)&=& [\exp\{\frac{k}{T_{\gamma}(t)(1+\Theta(\mathbf{k},\mathbf{x},t))}\}-1]^{-1}\nonumber \\
  &\simeq& f_\gamma^0+k\frac{\partial f_\gamma^0}{\partial k}\Theta(\mathbf{k},\mathbf{x},t),\label{theta}
\end{eqnarray}
where $f_\gamma^0=[\exp\{\frac{k}{T_{\gamma}}\}-1]^{-1}$ and $\Theta=\delta T/T$.
Here the zero-order temperature $T_\gamma(t)$ is a function of time only,
not space. The perturbation to the distribution function is characterized by $\Theta(\mathbf{k},\mathbf{x},t)$.
In the smooth zero-order universe, photons are
distributed homogeneously, that is, $T_\gamma$ is independent of $\vec{x}$ and isotropically, so $T_\gamma$ is
independent of the direction of propagation $\mathbf{k}$. We decomposed the perturbation $\Theta$ into a sum over Legendre polynomials,
\begin{equation}\label{theta1}
   \Theta(\mathbf{x},\mathbf{k},t)=\sum_l\Theta_l(k)\mathcal{P}_l(\mu),
\end{equation}
where $\mu$ is the dot product of the wave vector $\mathbf{k}$ and the direction of propagation \cite{MC} where in conjugate coordinate
\begin{equation}\label{theta1'}
   \Theta(\mathbf{K},\mathbf{k},t)=\sum_l\Theta_l(k)\mathcal{P}_l(\mathbf{K}.\mathbf{k}).
\end{equation}
By using Eqs.~(\ref{theta},\ref{theta1'}), we can expand the intensity of the CMB radiation as
\begin{equation}\label{theta02}
   I(\mathbf{K},\mathbf{k},t)\simeq I_0(k,t)+4k\frac{\partial I_0}{\partial k}\Delta I(\mathbf{K},\mathbf{k},t)+...\simeq=I_0(k,t)(1+ 4\Theta(\mathbf{K},\mathbf{k},t)),
\end{equation}
where $\Delta I(\mathbf{K},\mathbf{k},t)$ depends on $\Theta(\mathbf{K},\mathbf{k},t)$, so this quantity can represent into a sum over Legendre polynomials like Eq.(\ref{theta1'}),
\begin{eqnarray}\label{theta1'}
   \Delta I(\mathbf{K},\mathbf{k},t)=\sum_l\Delta I_l(k)\mathcal{P}_l(\mathbf{K}.\mathbf{k}).
\end{eqnarray}
On the other hand, at first glance, Compton scattering is a perfect mechanism for producing polarized
radiation. But to produce polarized radiation, the incoming radiation must have a nonzero
quadruple component \cite{cosowsky1994,MC}.
\begin{eqnarray}
    \frac{d}{dt}\Delta Q(\mathbf{K},\mathbf{k},t)&\approx& \sigma_T \bar{n}_e \Delta I_2(\mathbf{K},k,t)\sin^2\theta'\cos2\phi',\nonumber\\
     \frac{d}{dt}\Delta U(\mathbf{K},\mathbf{k},t)&\approx& -\sigma_T \bar{n}_e \Delta I_2(\mathbf{K},k,t) \sin^2\theta'\sin2\phi'\label{qu0}
\end{eqnarray}
where $\theta'$ and $\phi'$ are determined in Fig.(\ref{cmb3}) and
\begin{eqnarray}
 \Delta Q(\mathbf{K},\mathbf{k},t)=\left(4k\frac{\partial I_0}{\partial k}\right)^{-1} Q(\mathbf{K},\mathbf{k},t),\,\,\,\,\,\Delta U(\mathbf{K},\mathbf{k},t)=\left(4k\frac{\partial I_0}{\partial k}\right)^{-1} U(\mathbf{K},\mathbf{k},t),\label{AA}
\end{eqnarray}
here $\Delta Q(\Delta U)$ are dimensionless quantities  which should be smaller than $\delta T/T$, $\bar{n}_e$ is the electron number density and $I_0(k,t)$ is unpolarized intensity of the CMB which depends on $T_\gamma^4$.
 By substituting equation (\ref{qu0}) into (\ref{v6}) and choosing the spherical coordinates for $\hat{\mathbf{k}}$ with axis in the $\mathbf{K}$ direction (as shown in Fig.(\ref{cmb3})), we obtain the reasonable estimation for $V$-mode in terms of $\Theta_2$ at present time $t_0$
\begin{eqnarray}
   \Delta V(\mathbf{K},k,t_0) &\approx&3\times 10^{-3}\left(\frac{T_{0,\gamma}}{k}\right)\left(\frac{m_\nu}{0.1eV}\right)\left(\frac{n^0_\nu}{112/cm^3}\right)\,\bar\eta_\nu\,\sigma_T H_0^{-1} \bar{n}_e\,\Delta I_2(\mathbf{K},k,t_0) \nonumber\\
  &\times&\int_0^{1000}\frac{dz}{(1+z)^2}\frac{F(z)}{H(z)/H_0}
  \int_{z}^{1000}\frac{dz'}{H(z)/H_0},\label{v8}
\end{eqnarray}
where  $\Delta V$ is defined as dimensionless quantity as well as  $\Delta Q(\Delta U)$.
Then by considering relevant values for quantities which appear in above equation and making  integrations, we obtain
 \begin{eqnarray}
    \Delta V(\mathbf{K},k,t_0) &\approx&
    30 \left(\frac{T_{0,\gamma}}{k}\right)\left(\frac{m_\nu}{0.1eV}\right)\left(\frac{n^0_\nu}{112/cm^3}\right)
    \left(\frac{\bar{n}_e}{0.1cm^{-3}}\right)\,\bar\eta_\nu\,\Delta I_2(\mathbf{K},k,t_0). \label{v9}
\end{eqnarray}
The above equation is given for each mode of $\mathbf{K}$, but we interest to measure the value of $V$-parameter in real space coordinate $\mathbf{x}$ (we have to make inverse Furrier transform of above equation) and then two point function $<\Delta V(\mathbf{x},k,t_0)\Delta V(\mathbf{x},k,t_0)>$ where
 \begin{equation}\label{v9x}
    \Delta V(\mathbf{x},k,t_0)=\int\frac{d^3K}{(2\pi)^3}\,e^{i\mathbf{K}.\mathbf{x} } \Delta V(\mathbf{K},k,t_0)
 \end{equation}
Before doing this transformation,
we introduced $\xi(\mathbf{K})$, which is a random variable used
to characterize the initial amplitude of each mode of $\mathbf{K}$. It has the following statistical property
\begin{equation}\label{ips}
    \left<\xi^{\ast}(\mathbf{K}')\xi(\mathbf{K})\right>=(2\pi)^3\delta(\mathbf{K}'-\mathbf{K})\,p_s(K)
\end{equation}
$p_s(K)$ is so called primordial scaler power spectrum (index $"s"$ shows scaler perturbations). As discussed in \cite{para}, this quantity can be described as
\begin{equation}\label{ips1}
   p_s(K) = p_s(K_{\ast})\left(\frac{K}{K_{\ast}}\right)^{n_s-1}
\end{equation}
where $p_s(K_{\ast})$ and $n_{\ast}$ are determined at pivot scale $K_{\ast}\simeq 0.05/Mpc$. In general $n_s$ and $p_s(K_{\ast})$ depend on the pivot scale. As shown in above equation, the simplest case, neglecting a possible tensor component, the initial conditions are characterized by only two parameters $n_s$ and $p_s(K_{\ast})$.  For simplicity, we consider scale-invariant (Harrison- Zel'dovich spectrum) case with $n_s\simeq1$ and $p_s(K_{\ast})\propto A_s\frac{2\pi^2}{K_{\ast}^3}$. However, in principle, other initial condition are also possible. The total value of two point correlation function of $\Delta V$-mode can be written as
\begin{eqnarray}
 C^V&=& <\Delta V(\mathbf{x},k,t_0)\Delta V(\mathbf{x},k,t_0)>\simeq \int\frac{d^3K}{(2\pi)^3}p_s(K) |\Delta V(\mathbf{K},k,t_0)|^2 \nonumber\\
  &\simeq& 10^{3}\bar\eta_\nu^2\left(\frac{n^0_\nu(\mathbf{x})}{112/cm^3}\right)^2
  \left(\frac{m_\nu/0.1eV}{k/T_{0,\gamma}}\right)^2
    \left(\frac{\bar{n}_e}{0.1cm^{-3}}\right)^2\,C^T_2, \label{tpf1}
\end{eqnarray}
where
\begin{equation}\label{c2}
    C^T_2=\int\frac{ dK}{4\pi^2}\,K^2\,p_s(K) \,|\Delta I_2(\mathbf{K},k,t_0)|^2,
\end{equation}
Here we obtain the two point correlation function of $\Delta V$-mode as function two point correlation function of the angular power spectrum of temperature fluctuations [for more details about  $C^T_l$, see \cite{cosowsky1994,zal,para}]. Finally from equation (\ref{pertur}) and (\ref{tpf1}), the value of $C^V$ approximately is given
\begin{eqnarray}
  C^V\simeq 10^{-6}\left(\frac{N_\nu}{3}\right)^2\left(\frac{\bar\eta_\nu}{10^{-5}}\right)^2\left(\frac{n^0_\nu(\mathbf{x})}{112/cm^3}\right)^2
  \left(\frac{m_\nu/0.1eV}{k/T_{0,\gamma}}\right)^2
    \left(\frac{\bar{n}_e}{0.1cm^{-3}}\right)^2\,C^T_2, \label{v10x}
\end{eqnarray}
where $N_\nu$ is the number of neutrino flavors.
 Here we estimate the maximum value of $C^V$ in terms of two point correlation function of the angular power spectrum of temperature fluctuations $C^T_2$
where the value of $C^T_2$ is about a thousand  micro Kelvin square [see for example \cite{zal,para}]. As a result, the maximum value of $C^V$
is about $10^{-6}$ of the quadruple component of the the temperature power spectrum $C_2^T$ or larger then   Nano-Kelvin square. Of course we can generalize the above calculation for other components.
We should emphasize that
 we consider the contribution of anti-neutrino photon scattering  too. The calculation shows that the contribution of  anti-neutrino photon scattering is the same as neutrino photon scattering with the same sing [see appendix \ref{anti}], so in above calculation $n^0_{\nu }\equiv n^0_{\nu }+n^0_{\bar{\nu} }\sim 112/cm^3$.

\section{ After the last scattering: }

By considering photon-neutrino scattering at last scattering surface, we discuss and calculate the generation of the CMB's circular polarization. But we must remind during the propagation
from the last scattering surface to us, CMB photons
encounter large-scale structures and undergo significant
changes due to effects related to structure formation \cite{Cooray:2002vb}. The polarization modifications may occur during propagating in this large structure formation. On the other hand the presence of thermal electrons and
large-scale diffuse synchrotron emission towards galaxy clusters suggest the presence of the large scale magnetic fields \cite{clark}.  The presence of this large scale magnetic filed causes some modifications on the circular polarization of CMB due to the compton scattering which is discussed in \cite{fc,khodam,gio1,gio2}.The generation of circular polarization in the process of transfer of CMB within a large scale magnetic field and structure, due to the presence of the electrons, is discussed in \cite{fc}. They show the V-mode is about $10^{-9}$ for $ \lambda= 1cm$, $z=1000$ and for length scale about 1 Mpc and
number density of electrons about $0.1$ per $cm^3$,
\begin{equation}\label{fc1}
   V(k)\propto 10^{-9} \frac{\bar{n}_e}{0.1cm^{-3}}(\frac{B}{10\mu G})^2(\frac{\lambda}{1cm})^3\frac{L}{1Mpc},
\end{equation}
where is smaller than the maximum value of V-mode due to photon-neutrino scattering discussed in past section.  As shown in Eq.~(\ref{v9}), the value of V-mode due to neutrino-photon has the linear dependence on the wavelength $\lambda=1/k^0$ unlike the cubic dependence of the result of \cite{fc}.

The electron-photon scattering generates the linear polarization of CMB from unpolarized CMB \cite{cosowsky1994} but this process doesn't give any contribution for the CMB's circular polarization in the absence  of magnetic field. The probability of the generation of circular polarization via compton scattering in presence of  magnetic field is discussed in \cite{khodam}. The maximum value of the V-mode is given by
\begin{equation}\label{khodam1}
     V(k)\propto 10^{5} \frac{\bar{n}_e}{0.1cm^{-3}}(\frac{B}{10\mu G})(\frac{\lambda}{1cm})^3\frac{L}{1Mpc}(\frac{T_e}{m_e})^3,
\end{equation}
where the maximum of $\frac{T_e}{m_e}$ is about $10^{-6}$.  By considering $\frac{T_e}{m_e}$, the maximum value of V-model is about $10^{-13}$ where is smaller than result is given in Eq.~(\ref{v9}). Also the above equation has the linear dependence on magnetic field (unlike the result of \cite{fc}) and the cubic dependence on the wavelength (like the result of \cite{fc}) while in case of photon-neutrino scattering (\ref{v9}), the linear dependence on the wavelength appears. Of course we can determine the exact value for $V$- mode in Eqs.~(\ref{v9}), (\ref{fc1}) and (\ref{khodam1}) and we can be sure that the contribution of photon-neutrino scattering $V$ is the dominant contribution  because there are unknown parameters in each equation such as the value of the  large scale magnetic field, $\delta T / T(\vec{x},\vec{k})$ as well as $I(\vec{x},\vec{k})$. But by considering  relevant values for the large scale magnetic field and anisotropic, we can find that the contribution of photon-neutrino scattering to generate CMB's circular polarization is dominant.

 Another effect of  the large-scale structure and magnetic field which needs to be discussed  may appear as  the circular polarization  convert to linear one due to photon-neutrino and Compton scattering, that means
 \begin{equation}\label{con}
   \frac{ dQ(U)}{dt}(\vec{k})\propto V(\vec{k})
 \end{equation}
First, we investigate the conversion of  circular polarization to linear one via Compton scattering. The Compton scattering in the absent of magnetic field doesn't give any term like (\ref{con}), see section $IV$ in \cite{cosowsky1994}. But the Compton scattering in the presence of magnetic field gives \cite{khodam}
 \begin{equation}\label{con1}
    Q(U)(\vec{k})\propto 10^{5} \frac{\bar{n}_e}{0.1cm^{-3}}(\frac{B}{10\mu G})(\frac{\lambda}{1cm})^3\frac{L}{1Mpc}(\frac{T_e}{m_e})^3 V(\vec{k})
 \end{equation}
If we substitute the value of the $ V(\vec{k})$ from equation (\ref{v6}) and the relevant value of parameters which appear in above equation, The maximum value of $Q(U)(\vec{k})$ becomes very small. So we can neglect the conversion of the circular to linear polarization via Compton scattering. To investigate the conversion of the circular to linear polarization via photon-neutrino scattering, we use Eqs.~(\ref{bo}) and (\ref{fws2}) which give
\begin{eqnarray}
  \frac{dQ(\mathbf{k})}{dt} &\approx& \frac{1}{6}\frac{1}{(4\pi)^2}\frac{e^2g_w^2}{M^2_W k^0}\int d\mathbf{q}\,\,n_\nu(x,q)[\bar{U}_{r}(q)(1+\gamma_5) \nonumber\\
  &\times& (\epsilon_{1}\!\!\!\!\!/\,\,q.\epsilon_1-\epsilon_{2}\!\!\!\!\!/\,\,q.\epsilon_2)U_r(q)]V(\mathbf{k}).\label{con2}
\end{eqnarray}
By substituting $V(\mathbf{k})$ from equation (\ref{v6}), the value of $Q(\mathbf{k})$ is proportional to $(G^F)^2$ which become very small. It is so negligible

\section{ Conclusion.}

In this letter, by approximately solving the first order of Quantum
Boltzmann Equation for the density matrix of a photon ensemble, and
time-evolution of Stokes parameters, we show that the linear polarizations of the CMB can convert to circular
polarizations by scattering the CMB photon on cosmic neutrinos background C$\nu$B. The maximum value of the $V$-Stokes parameter in $\mathbf{K}$ direction is given by (\ref{v9}) at frequencies of a few $GHz$ with the  linear dependence  on the wavelength and  the C$\nu$B perturbations  of  pressure $\delta P_\nu$ and shear stress $\sigma_\nu$. To have a measurable quantity for circular polarization, we calculate  $C^V=<VV>$. By considering the average value of $\bar\eta_\nu$ about the fluctuation temperature $\bar\eta_\nu\leq\delta T/T\big|_\nu$, the maximum value of $C^V$ is about $10^{-6}$ of $C^T_2$ angular power spectra or lager than Nano-kelvin square. We should mention, we only try to estimate analytically the value of circular polarization due to CMB and C$\nu$B scattering. But to have the exactly value, we should numerically solve Boltzmann equation for CMB and C$\nu$B in during a model for expansion universe  which leave it here.

It is expected that the polarization data, which will become available with the Plank 2014 data, provide valuable information on the nature of the CMB anomalies (with resolution in range of Nano-kelvin) \cite{plank} and  there are also other
high resolution polarization experiments such as $\rm{ACTPol}$
\cite{act}, $\rm{PIXIE}$ \cite{pix}, $\rm{SPIDER}$ \cite{spider}. Of course we can not exactly compare the effects of photon-neutrino scattering on the circular polarization of the CMB with other interactions without any knowledge about the pattern and distribution of the initial linear polarization but we can discuss about its maximum.
Our value for circular polarization (\ref{v9}) is lager than the one which is given by \cite{fc} and comparable with the bound reported in \cite{nature}. In the work reported in \cite{fc}, they show that Faraday conversion process during the propagation of polarized CMB
photons through regions of the large-scale structure containing magnetized relativistic plasma, such
as galaxy clusters, will lead to a circularly polarized contribution of order $10^{-9}$ at frequencies of $10 GHz$ with the cubic dependence on the wavelength (our result has the  linear dependence  on the wavelength) and square dependence on the large scale magnetic field.
Refs.~\cite{gio1,gio2} have argued that the presence of a large-scale magnetic field prior
to equality can affect the photon-electron and the photon-ion scattering, this leads to the radiation becomes circularly polarized and the induced $VV$ angular power spectra have been computed. Their results are comparable with the result of \cite{fc}. In \cite{khodam}, the effect of the large-scale of magnetic field on the compton scattering has been discussed which leads to generate circular polarization for CMB. The band on $V$-mode reported in \cite{khodam} is very smaller than our result and has the cubic dependence on the wavelength (our result has the  linear dependence  on the wavelength), the linear dependence on the large scale magnetic field and the cubic dependence on $T_e/m_e$. The band on $V$-mode reported in \cite{fc} can be larger than our result if we have the large-scale of magnetic field in order of $B>10mG$. Also in  \cite{khodam}, they show that CMB polarization acquires a small degree of circular polarization
when  the quantum electrodynamic
sector of standard model is extended by Lorentz non-invariant operators as well as
non-commutativity theory. These results contain  Lorentz non-invariant and non-commutativity parameters which we don't know the exactly values of them.
In Ref.~\cite{xue}, it has been shown that circular polarizations of radiation fields can be generated
from the effective Euler-Heisenberg Lagrangian in order of $10^{-10}$K which is very small.
The transformation plane of the
polarization into circular polarization via photon-photon interactions mediated by the neutral hydrogen background, $\gamma+\gamma+atom\rightarrow \gamma+\gamma+atom$, through completely forward processes, has been discussed in \cite{pp}. The ratio of circular to plane
polarization intensities $V/Q$ is predicted to be at the level of several times $10^{-5}$ for some regions of angular size less than $1/300$ and with large plane polarizations. So the value of the circular polarization (from CMB and C$\nu$B forward scattering)
seems to be  large enough to detect. And the other hand,  as we already mentioned that the detection of C$\nu$B is hardly possible due to the weak interaction of neutrinos with matter and due to their low energy, however  the measuring of CMB's circular polarization may give us  the good experimental testifier for cosmic neutrino back ground C$\nu$B.

\vskip0.5cm
\noindent
{\bf Acknowledgment.}
\hskip0.1cm
I would like to thank
S. S. Xue and  I. Motie  for fruitful discussion.


\newpage
\section{Appendix: Anti-neutrino-photon scattering}\label{anti}
\begin{figure}
\begin{center}
  \includegraphics[width=0.5\columnwidth]{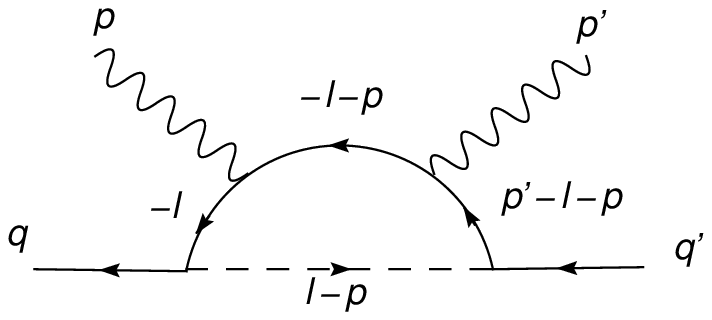}\\
  \caption{The typical diagram of photon-antineutrino scattering is given in this plot. }\label{cmb2}
  \end{center}
\end{figure}

The first order of photon - anti-neutrino  Hamiltonian interaction is given
 \begin{eqnarray}
   H^0_I &=& \int d\mathbf{q} d\mathbf{q'} d\mathbf{p} d\mathbf{p'} (2\pi)^3\delta^3(\mathbf{q'} +\mathbf{p'} -\mathbf{p} -\mathbf{q} ) \nonumber \\
    &\times& \exp[it(q'^0+p'^0-q^0-p^0)]\left(d_{r}a^{\dagger}_{s'}(\mathcal{M}'_1+\mathcal{M}'_2)a_s\,d^{\dagger}_{r'}\right)
 \end{eqnarray}

where the amplitude of the diagram shown in  Fig.(\ref{cmb2}) and its crossing is given by
 \begin{eqnarray}
  \mathcal{M}'_1+\mathcal{M}'_2 &=& \frac{1}{8}e^2g_w^2\int\frac{d^4l}{(2\pi)^4}D_{\alpha\beta}(l-q)\bar{\mathcal{V}}_{r}(q)\gamma^\alpha (1-\gamma_5)S_F(p'-p-l)\nonumber\\
   &\times& \left[\ep_{s}\,S_F(-l-p)\,\ep_{s'}\,+\,\ep_{s'}\,S_F(p'-l)\,\ep_{s'}\,\right]
   S_F(-l)\gamma^\beta (1-\gamma_5)\mathcal{V}_{r'}(q'),
\end{eqnarray}
Next we substitute above equation into (\ref{fws}) then we make average of this exsection value of  anti-neutrino-photon forward scattering
(similar to equation (\ref{fws1}) for photon neutrino scattering) and then we have
\begin{eqnarray}
  i\langle[H^0_I, D^0_{ij}({\bf k})]\rangle &=& +\frac{i}{16}e^2g_w^2\int d\mathbf{q} (\rho_{s'j}({\bf k})\delta_{is} -\rho_{is}({\bf k})\delta_{js'})n_{\bar{\nu}}(x,q)\nonumber\\
  &\times& \int\frac{d^4l}{(2\pi)^4}D_{\alpha\beta}(-q+l)\bar{\mathcal{V}}_{r}(q)\gamma^\alpha (1-\gamma_5)S_F(-l)\nonumber\\
   &\times& \left[\,\ep_{s}\,S_F(-l-k)\,\ep_{s'}\,+\,\ep_{s'}\,S_F(-l+k)\,\ep_{s}\,\,\right]
   S_F(-l)\gamma^\beta (1-\gamma_5)\mathcal{V}_r(q).\label{fws1a}
\end{eqnarray}
By helping Dimensional regularization and Feynman parameters, we will go forward to obtain  the leading order term of the right side of the above equation, then

\begin{eqnarray}
  i\langle[H^0_I, D^0_{ij}({\bf k})]\rangle &=& -\frac{1}{16}\frac{1}{4\pi^2}e^2g_w^2\int d\mathbf{q} (\rho_{s'j}({\bf k})\delta_{is} -\rho_{is}({\bf k})\delta_{js'})n_{\bar{\nu}}(x,q)\nonumber\\
  &\times& \int_0^1dy\int_0^{1-y}dz\frac{(1-y-z)}{zM^2_W}\bar{\mathcal{V}}_{r}(q)(1+\gamma_5)(2z q\!\!\!/\epsilon_{s'}.\epsilon_s\nonumber\\
   &+& \,2z(\,\ep_{s'}\,\mathbf{q}.\epsilon_s\,+\ep_{s}\,\mathbf{q}.\epsilon_{s'}\,)-
   (3y-1)k\!\!\!/\,(\ep_{s}\,\ep_{s'}\,-\,\ep_{s'}\,\ep_{s}\,)\,)
  \mathcal{V}_r(q).\label{fws2a}
\end{eqnarray}
By use the gamma-matrix identity, the polarization vector properties $k.\epsilon_i=0$ and
$\epsilon_i.\epsilon_j=-\delta_{ij}$. Finally  $dV/dt$ is given as following
\begin{eqnarray}
  \frac{dV(\mathbf{x},\mathbf{k})}{dt} &\approx& +\frac{1}{6}\frac{1}{(4\pi)^2}\frac{e^2g_w^2}{M^2_W k^0}\int d\mathbf{q}\,\,n_{\bar{\nu}}(x,q)\bar{\mathcal{V}}_{r}(q)(1+\gamma_5) \nonumber\\
  &\times& \left[(\ep_{1}\,\,q.\epsilon_1\,-\,\ep_{2}\,q.\epsilon_2)Q(\mathbf{x},\mathbf{k})-
  (\ep_{1}\,q.\epsilon_2+\ep_{2}\,q.\epsilon_1)U(\mathbf{x},\mathbf{k})\right]\mathcal{V}_r(q),\label{v1a}
\end{eqnarray}
And also  belove equations are useful:
\begin{equation}\label{sn2a}
   \bar{\mathcal{V}}_r(q)\gamma^\mu \mathcal{V}_s(q)=2q^\mu\delta^{rs},\,\,\,\,\, \frac{1}{2}\sum_r\bar{\mathcal{V}}_r(q)\gamma^\mu(1\pm\gamma^5) \mathcal{V}_r(q)=2q^\mu.
\end{equation}

By substituting above equation into (\ref{v1a}) and assuming $\nu_{\bar\nu}=n_{\nu}$, we arrive to
\begin{eqnarray}\label{v3a}
  \frac{dV(\mathbf{x},\mathbf{k})}{dt}\big|_{\bar\nu} &=& \frac{dV(\mathbf{x},\mathbf{k})}{dt}\big|_\nu
\end{eqnarray}
This means that the anti-neutrino photon scattering affects the circular polarization of CMB as same as the neutrino photon scattering with the same value and sing.
\end{document}